\documentclass[a4paper, oneside, twocolumn, notitlepage, 10pt]{style/extarticle_ecoc2015}
\usepackage{style/ecoc2015}
\usepackage{color}
\usepackage{tumcolor}
\usepackage{tikz}
\usepackage{pgfplots}
\usetikzlibrary{backgrounds,arrows}
\usepackage{siunitx}
\usepackage{subfig}

\newsavebox\myinsetbox

\pgfkeys{/tikz/.cd,
  background color/.initial=graphicbackground,
  background color/.get=\backcol,
  background color/.store in=\backcol,
}
\tikzset{background rectangle/.style={
    fill=\backcol,
  },
  use background/.style={    
    show background rectangle
  }
}

\newcommand\blfootnote[1]{%
  \begingroup
  \renewcommand\thefootnote{}\footnote{#1}%
  \addtocounter{footnote}{-1}%
  \endgroup
}

\begin{document}
\selectlanguage{american}    % Standard Language

%\title{Experimental Demonstration of Probabilistically Shaped 64-QAM with a Simple Rate-Adaptation Scheme Based on a Single FEC Decoder}
\title{Experimental Demonstration of Capacity Increase and Rate-Adaptation by Probabilistically Shaped 64-QAM\vspace*{-1.5ex}}

%------------------------------------------------- Authors-----------------------------------------------------%

\author{
    F. Buchali\textsuperscript{(1)}, G. B\"ocherer\textsuperscript{(2)}, 
    W. Idler\textsuperscript{(1)}, L. Schmalen\textsuperscript{(1)}, P. Schulte\textsuperscript{(2)}, and F. Steiner\textsuperscript{(2)}
\vspace*{-1.5ex}}

\maketitle                  % Create title and author

%------------------------------------------ Description of Authors ----------------------------------------------%

\begin{strip}
 \begin{author_descr}

   \textsuperscript{(1)} Alcatel-Lucent Bell Labs, Stuttgart, Germany,
   \uline{fred.buchali@alcatel-lucent.com} 

   \textsuperscript{(2)} Technische Universit\"at M\"unchen, Germany,
   \uline{georg.boecherer@tum.de} 
 \end{author_descr}
\end{strip}

\setstretch{1.1}

%-------------------------------------------------- Abstract ---------------------------------------------------------%

\begin{strip}
  \begin{ecoc_abstract}
    % NOTE: Don't use a blank line here but start abstract right away to avoid an extra line break
%    \textcolor{blue}{Your 40 words abstract, which will appear in the conference program, should be an explicit summary of the paper that states the problem, the methods used as well as the major results and conclusions. It should be complementary to rather than a repeat of the title.}
 %   We demonstrate in an experiment that probabilistic shaping increases the capacity of the optical fiber and that it can be used to realize a flexible, versatile transmission system that can operate at different net data rates at a fixed constellation and FEC overhead.
 %   Optical transmission experiments are used to demonstrate that probabilistic shaping increases optical fiber capacity. Shaping further enables the realization of a flexible transmission system operating at different data rates with fixed bandwidth, constellation and overhead.
 We implemented a flexible transmission system operating at adjustable data rate and fixed bandwidth, baudrate, constellation and overhead using probabilistic shaping. We demonstrated in a transmission experiment up to 15\% capacity and 43\% reach increase versus 200 Gbit/s 16-QAM. 
\end{ecoc_abstract}
\end{strip}

\vspace*{-5.5ex}
\section{Introduction}
Future optical metro and long-haul networks require transceivers that maximize spectral efficiency and throughput and that optimally exploit all available resources. For example, a transceiver that operates on a short network segment with high {signal-to-noise ratio} (SNR) should achieve a high spectral efficiency to maximize the net data rate over this segment. Similarly, a transceiver operating on a long network segment (e.g., an intercontinental route) with low OSNR should use either a lower order modulation format or a {forward error correction} (FEC) code with high overhead to ensure reliability. 
\blfootnote{The work of G. B\"ocherer and P. Schulte was supported by the German Federal Ministry of Education and Research in the framework of an Alexander von Humboldt Professorship. F. Steiner was supported by Technische Universit\"at M\"unchen, Institute of Advanced Studies (IAS). F. Buchali and L. Schmalen were supported by the German BMBF in the scope of the CELTIC+ project SASER/SaveNet.}

Today's coherent optical transceivers typically use a handful of coding and modulation modes for flexibility, e.g., different modulation formats and one or two FEC engines with different overheads. The flexibility of such systems is limited because they are only coarsely adaptable. The different operating modes often require changes in the baudrate or the FEC overhead, which poses implementation problems. Furthermore, conventional coded modulation schemes 
%, often based on the pragmatic {bit-interleaved coded modulation} (BICM), 
 show a gap to Shannon capacity that can be overcome only by using modulation formats that have a Gaussian-like shape~\cite{refDarISIT14,refSmith12,refYankov14}. Such shaping is known to improve the non-linear tolerance as well\cite{refDarISIT14}.

\begin{figure}[t!]
	\vspace*{-3ex}
	\footnotesize
	\centering
	\subfloat[$H(P_1) = \num{5.73}$ bits]{\includegraphics[scale=0.32]{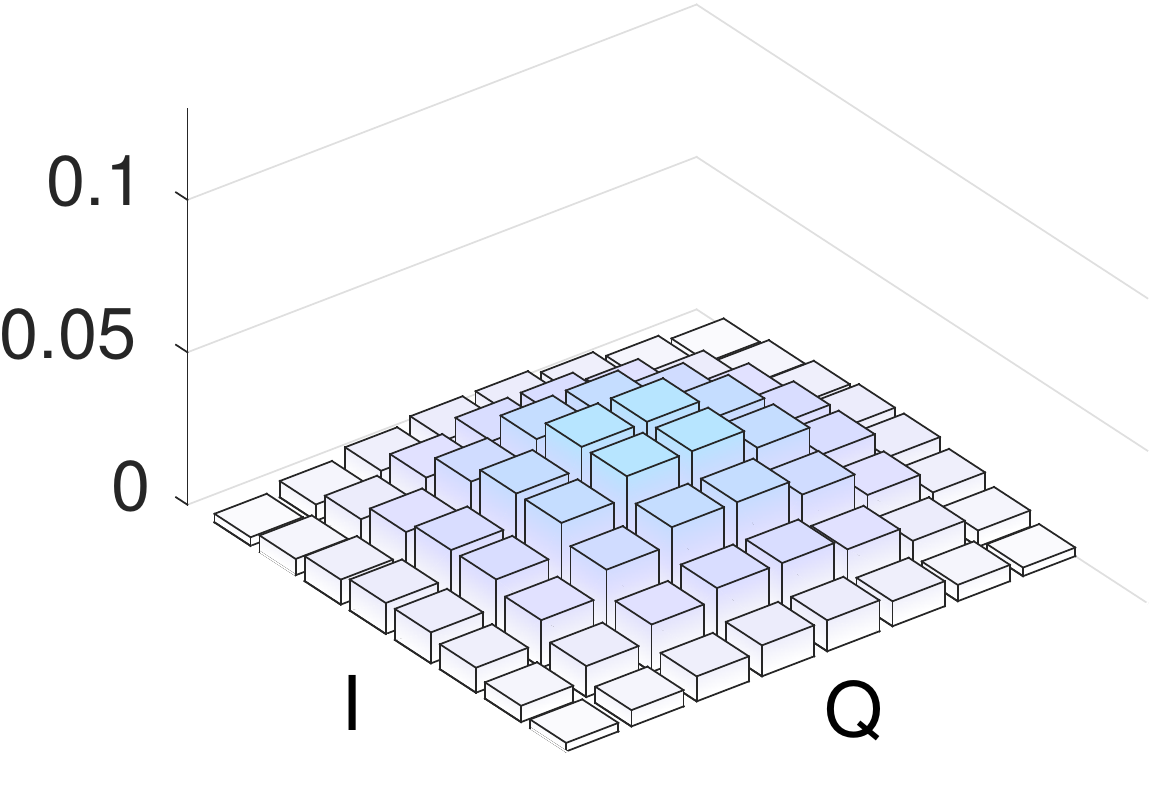}}
	\subfloat[$H(P_2) = \num{5.23}$ bits]{\includegraphics[scale=0.32]{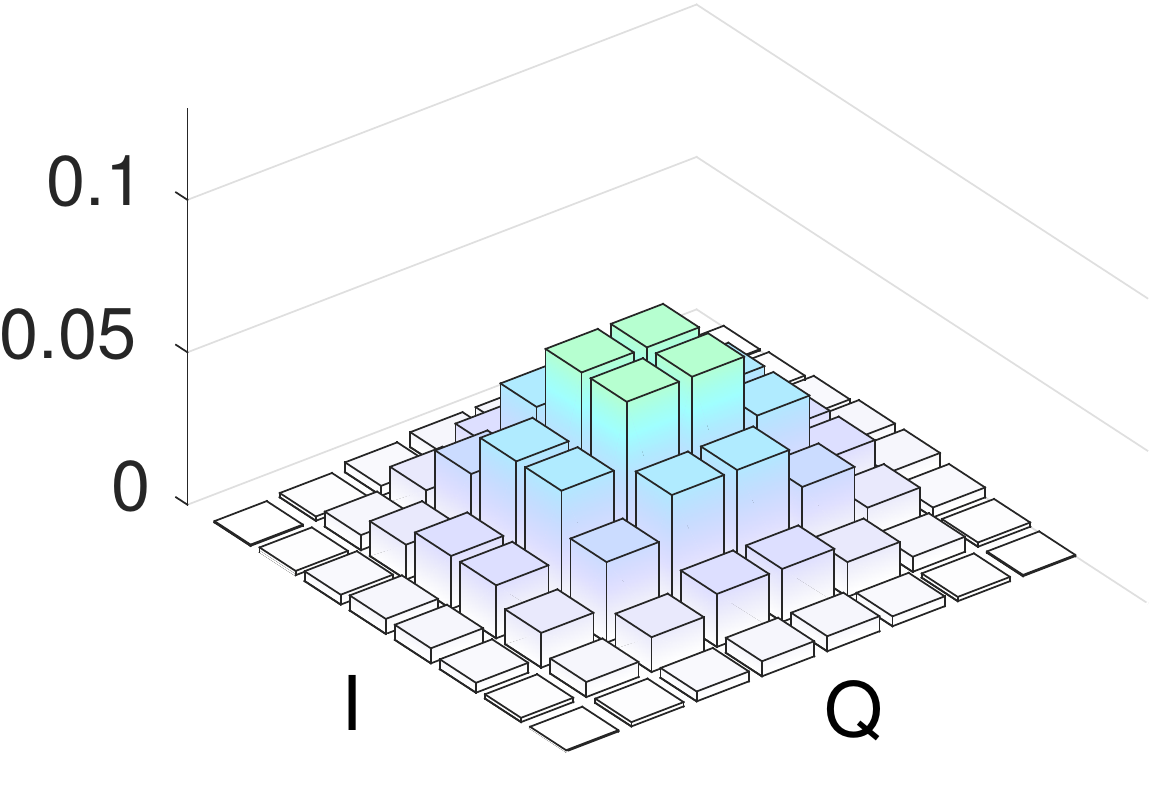}}\\\vspace*{-3ex}
	\subfloat[$H(P_3) = \num{4.60}$ bits]{\includegraphics[scale=0.32]{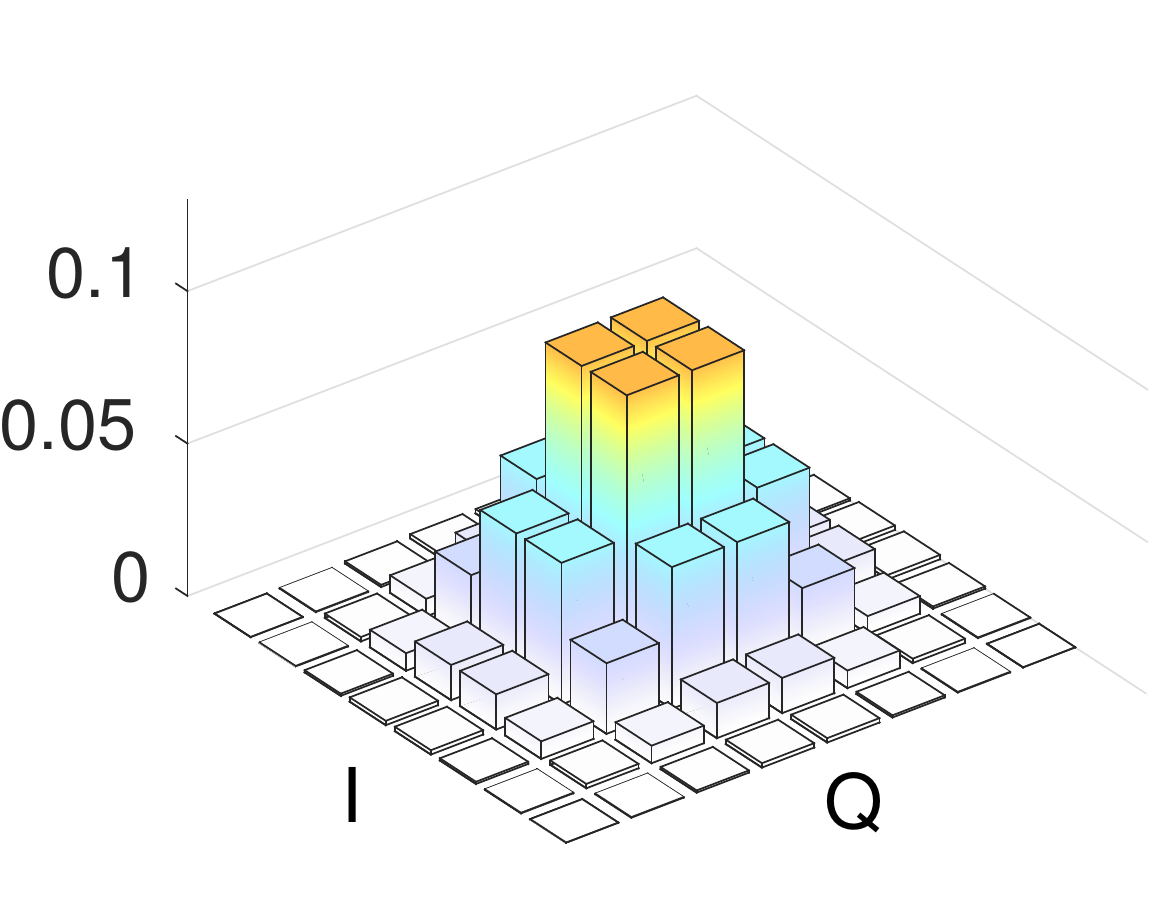}}
	\subfloat[$H(P_4) = \num{4.13}$ bits]{\includegraphics[scale=0.32]{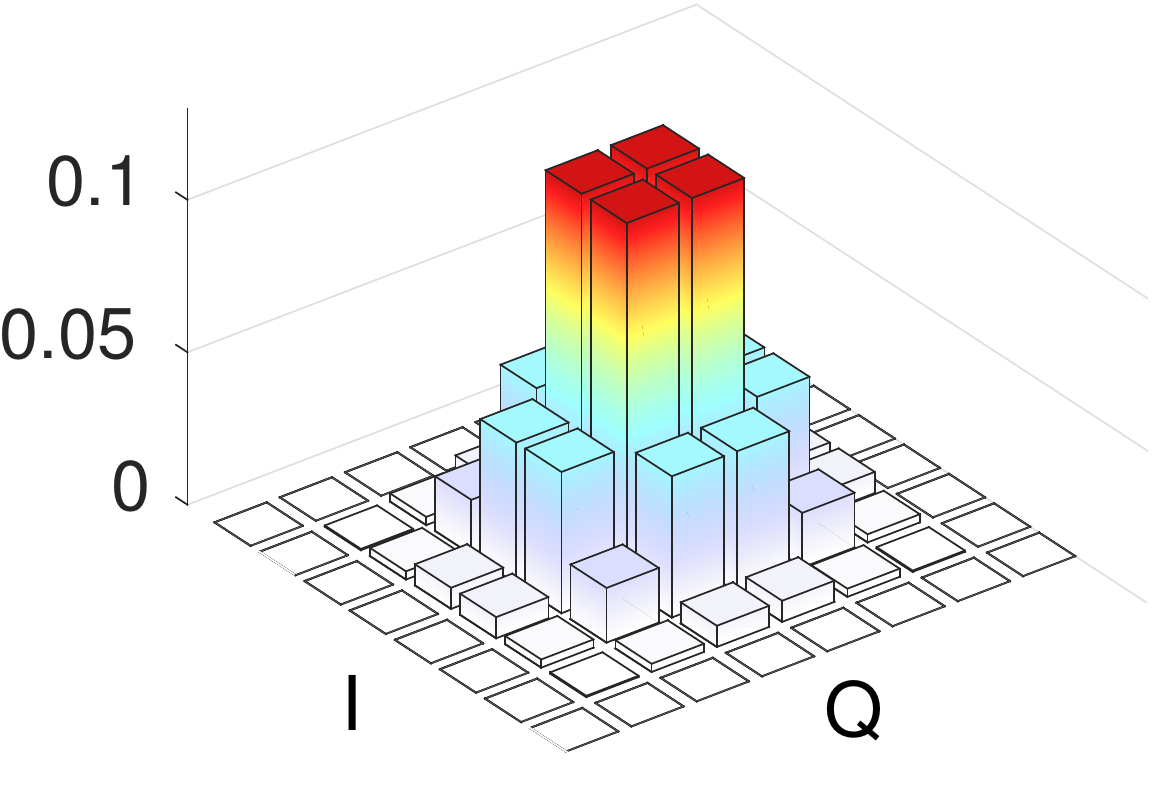}}
	\caption{Graphical illustration of the four employed probability distributions for PS-64-QAM. The bars indicate the probability of each modulation symbol. From (a) to (d), the distributions become more shaped and the entropies $H(P_i)$ decrease.}
	\label{fig:fig_distributions1}
\end{figure}

In this paper, we propose a system that uses \emph{probabilistic constellation shaping} (PS) to close the gap to capacity. The system design has an unprecedented flexibility in terms of transmission rate \emph{without} increasing the system and implementation complexity. For the first time, we experimentally verify a coded modulation scheme with rate adaptation~\cite{refGB15} that substantially increases the transmission distance and that is flexible even though it uses fixed FEC overhead, fixed modulation format and fixed baudrate and bandwidth. The key step is to introduce a \emph{distribution matcher}\cite{refPS15} (DM) that generates a non-uniform modulation symbol sequence (see Fig.~\ref{fig:fig_distributions1}) from the data sequence. We find that the gains predicted by theory and simulations can be achieved with a practical, low-complexity system.

\section{Rate-Adaptive Constellation Shaping}
\vspace*{-1.5ex}
\begin{figure}[tbh!]
\footnotesize
\centering
\includegraphics{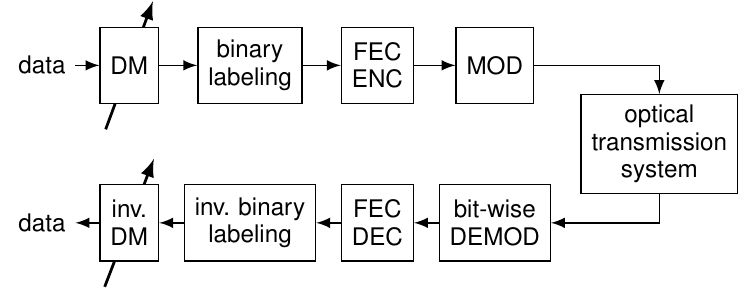}
\vspace*{-2ex}
\caption{System model of coding and modulation.}
\label{fig:system_model}
\end{figure}
State-of-the-art systems assume that all symbols are used with equal probability. We use the rate adaptive coding and modulation scheme proposed in\cite{refGB15} which deliberately assigns different probabilities to each modulation symbol. Figure~\ref{fig:system_model} shows the high-level model. The key device is the DM\cite{refPS15,CCDM} that transforms the sequence of data bits into a sequence of non-uniformly distributed (shaped) symbols. The shaped symbols are represented by binary labels and encoded by a binary FEC encoder, which is systematic to preserve the distribution of the shaped symbols. The FEC encoder output is mapped to a sequence of complex {quadrature amplitude modulation} (QAM) symbols. This sequence is fed to the optical transmission system, which outputs a noisy sequence of complex QAM symbols. The demodulater uses the noisy sequence to calculate bit-wise {log-likelihood ratios} (LLRs) that are fed to the FEC decoder. The decoded symbols are transformed back to the data bits by an inverse DM. 

As our binary FEC code, we use a {spatially coupled} rate 5/6 LDPC (SC-LDPC) code~\cite{SchmalenSC}. In principle, we could use any FEC scheme, but we opt for SC-LDPC codes because of their excellent performance. The SC-LDPC code has no error floor and allows for soft decision decoding. SC-LDPC codes are robust and show the same error performance in different scenarios. In particular, our code has the same error performance when operated with different constellation distributions.

Let $P$ denote the constellation distribution after modulation imposed by the DM, let $c$ denote the code rate of the FEC code, and let $m=6$ be the number of bit-levels of 64-QAM. The transmission rate is given by\vspace*{-2ex}
\begin{align}
R=H(P)-(1-c)\cdot m\;\left[\frac{\text{\footnotesize bits}}{\text{\footnotesize QAM\, symbol}}\right]\label{eq:transmission rate}\\[-4ex]\nonumber
\end{align}
where $H(P)$ is the entropy of $P$ in bits\cite{refGB15}. By \eqref{eq:transmission rate}, we can transmit at different rates $R$ by changing the distribution $P$ and using the same FEC code. Following~\cite{refFehenbergerOFC15}, we choose $P$ from the family of Maxwell-Boltzmann distributions, see Fig.~\ref{fig:fig_distributions1} for an illustration of the four distributions $P_1, P_2, P_3, P_4$ and the resulting probabilistically shaped PS-64-QAM constellations that we use in our experiment. The corresponding entropies $H(P_i	)$ are listed in the captions of Fig.~\ref{fig:fig_distributions1}. 
\vspace*{-1ex}

%For 64-QAM and code rate $c=5/6$, the transmission rate in bits/QAM symbol evaluates to
%\begin{align}
%&R_1=H(P_1)-(1-5/6)\cdot6=4.7253\label{eq:p1}\\
%&R_2=H(P_2)-(1-5/6)\cdot6=4.2253\\
%&R_3=H(P_3)-(1-5/6)\cdot6=3.6039.\label{eq:p3}
%\end{align}

%-------------------------------------------------- Section 4 -------------------------------------------------------%

\section{Experimental Setup}
\begin{figure}[t!]
 \centering
\includegraphics{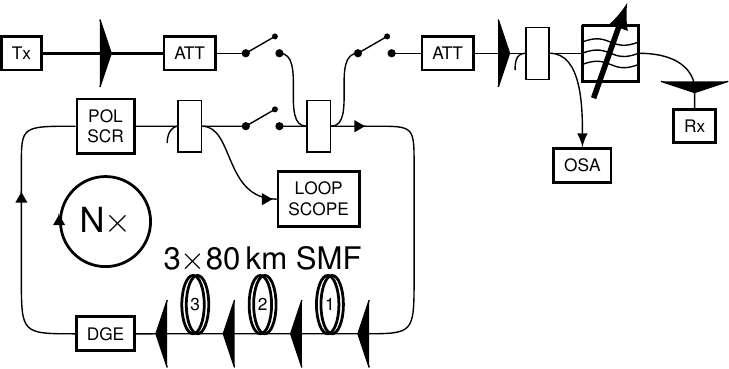}
 \caption{Illustration of the experimental setup.}
 \vspace*{-0.3ex}
 \label{fig:experimental_setup}
\end{figure}

Optical transmission experiments have been conducted using the standard coherent transmission loop setup shown in Fig.~\ref{fig:experimental_setup}. The transmitter is based on an 88 GSamples/s quad-{digital-to-analog converter} (DAC) and a linear amplifier driving the dual polarization IQ-modulator. The channel under test was operated at 32 GBaud. In the transmitter (Tx) DSP we incorporated Nyquist filtering with \num{0.15} roll-off factor and a pre-emphasis to compensate for the bandwidth limitations of the DAC and driver amplifier. The precalculated sequences are loaded into the memory of the DAC and transmitted periodically. In addition to the channel under test, we used $2\times 4$ load channels operated at 32 Gbaud DP-QPSK with 4 nm guard bands. The loop consists of three \SI{80}{\kilo\meter} SMF spans. The signals are amplified in single stage EDFAs with a noise figure of \SI{5}{\decibel}. The launch power was optimized individually for all experiments. A standard dual-polarization coherent receiver was used with high bandwidth differential photodiodes and \SI{33}{\giga\hertz} bandwidth AD conversion at 80 Gsamples/s. We stored sequences with \num{500000} samples and processed the data offline. 
We applied two receiver {digital signal processors} (DSPs) for data processing: The regular QAM modes and the PS-64-QAM mode with distribution $P_1$ were processed using a standard DSP with blind adaptation whereas the $P_2$, $P_3$ and $P_4$ PS-64-QAM modes were processed using a data aided approach.
The blind adaptation DSP includes re-sampling to 2 sample/symbol, {chromatic dispersion} (CD) compensation, polarization de-multiplexing using a butterfly equalizer with a simple {constant modulus algorithm} (CMA) for adaptation, frequency offset compensation and \mbox{4th}-power phase estimation. After DSP, we include demodulation and soft-decision decoding as detailed in~\cite{refGB15}. Note that using the data-aided DSP for regular QAM and PS-64-QAM ($P_1$) modes does not lead to noteworthy performance improvements.\vspace*{-1ex}
\begin{figure*}[t!]
\footnotesize
\centering
\includegraphics{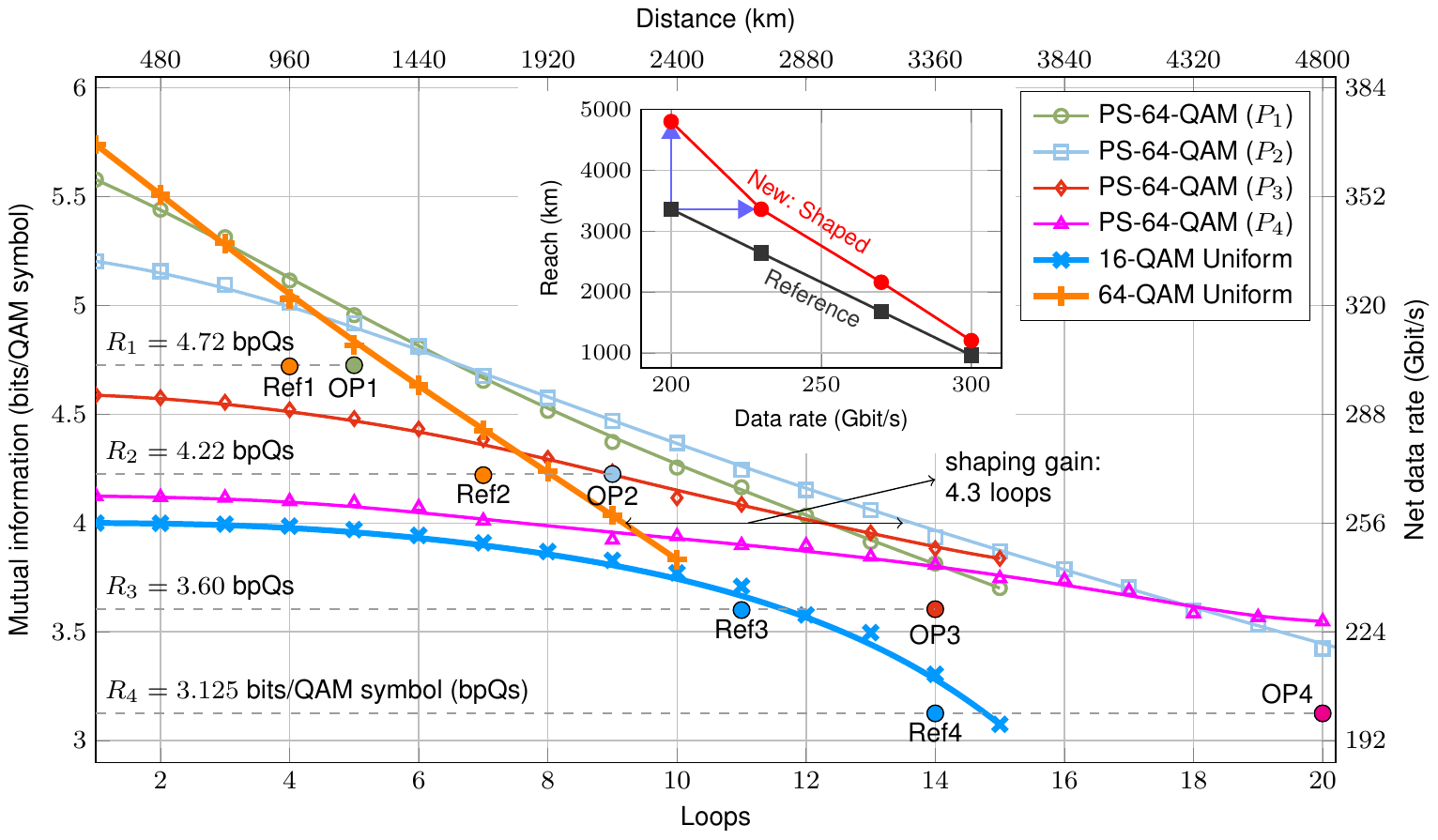}
%\vspace*{-2ex}
\caption{Experimentally measured mutual information for the regular uniform distribution and PS-64-QAM with the four shaped distributions $P_1$,  $P_2$, $P_3$ and $P_4$. A shaping gain of \num{4.3} loops, i.e., \SI{1032}{\kilo\meter}, can be observed at a mutual information of 4 bits per QAM symbol. From Table~\ref{tab:results}, we display the reference operating points Ref$i$ and the operating points OP$i$.}
\label{fig:fig_shaping_gain}
\end{figure*}

\section{Results}

Fig.~\ref{fig:fig_shaping_gain} shows the results of the transmission experiment. We display the measured {mutual information} (MI), which give the maximum achievable rate assuming ideal FEC(s)\cite{AlvaradoMI}. For practical FECs, we must add a penalty that depends on the actual FEC realization. 

We fix the FEC code with overhead 20\% ($c=5/6$) and use the four distributions in Fig.~\ref{fig:fig_distributions1}. The data rates are given by~\eqref{eq:transmission rate} and the respective entropies are shown in the captions of Fig.~\ref{fig:fig_distributions1}. The reach is determined by the maximum distance where we observe error-free decoding.  The rates and achievable distances are summarized in Tab.~\ref{tab:results}. Note that all systems we compare use the same channel bandwidth and we trade off reach against net data rate. As a reference system, we consider 64-QAM and 16-QAM with uniformly distributed symbols and varying FEC overheads. The points OP$i$ show how one can trade off reach against net data rate solely by adapting the distribution via the DM. At \num{300}\,Gbit/s we found \num{25}\% reach increase rising up to \num{43}\% at \num{200}\,Gbit/s (inset of Fig.~\ref{fig:fig_shaping_gain}). The reach increase is inline with the theoretically expected gains~\cite{refGB15,refDarISIT14}. Probabilistic shaping increases reach for a fixed net data rate and yields a higher net data rate for a fixed distance as compared to conventional systems. 

\begin{table}[tbh!]

\caption{Summary of the main results}\label{tab:results}
\begin{small}
\begin{tabular}{@{}c@{\!}ccc@{\!}c@{}}
& Net data & FEC & Constel- &Achievable \\
& rate (Gbit/s) & OH &lation & dist. (km) \\
\hline
Ref1 & 300 & 28\% & 64-QAM & 960 \\
Ref2 & 270 & 43\% & 64-QAM & 1680 \\
Ref3 & 230 & 11\% & 16-QAM & 2640 \\
Ref4 & 200 & 28\% & 16-QAM & 3360 \\
\hline
OP1 & 300 & 20\% & PS-64-QAM ($P_1$) & 1200 \\
OP2 & 270 & 20\% & PS-64-QAM ($P_2$) & 2160 \\
OP3 & 230 & 20\% & PS-64-QAM ($P_3$) & 3360 \\
OP4 & 200 & 20\% & PS-64-QAM ($P_4$) & 4800  
\end{tabular}
\end{small}
\vspace*{-2ex}
\end{table}

 %-------------------------------------------------- Section 5 -------------------------------------------------------%

\section{Conclusions}

We demonstrated the first optical transmission experiment of probabilistically shaped higher order constellation schemes. The results show a 15\% capacity increase and 43\% reach increase versus \num{200}\,Gbit/s 16-QAM. It realizes simple rate adaptation by adjustable shaping that allows to select arbitrary operating points without changing FEC overhead, constellation, and symbol rate. 

\vspace{-0.2cm}

%-------------------------------------------------- Section 9 -------------------------------------------------------%
%-------------------------------------------------- Literature -------------------------------------------------------%
\bibliographystyle{abbrv}
\begin{spacing}{1.35}

\end{spacing}
\vspace{-4mm}

%%%%%%%%%%%%%%%%%%%%%%%%%%%%%%%%%%%%%%%%%%%%%
%---------------------------------------------- End of Document -----------------------------------------------%
\end{document}